\documentclass[prd,twocolumn, tightenlines, superscriptaddress, 10pt]{revtex4-1}
\usepackage{bm}
\usepackage{xcolor}
\usepackage[
left=2.00cm,
right=2.00cm,
top=2.00cm,
bottom=2.00cm
]{geometry}
\usepackage{amsmath}
\begin{document}

\title{Comment on ``Basic observables for the accelerated electron and its field"}

\author{V. Hnizdo}
\email{hnizdo2044@gmail.com}
\affiliation{2044 Georgian Lane, Morgantown, WV 26508, USA}
\author{G. Vaman}
\email{vaman@ifin.nipne.ro}
\affiliation{Institute of Atomic Physics, P.\,O.\,Box MG-6, Bucharest, Romania}
 
\begin{abstract}
We show that the ``defective" terms in the expression that Dondera [Phys.\,Rev.\,D {\bf 98}, 096008 (2018)] obtained for the momentum of the retarded field of an accelerating point charge are mathematically well justified. The repair should not be sought in assigning an accelerating ``bare" charge {\it ad hoc} compensating attributes. We advance a conjecture, supported by  published work, concerning the Hadamard finite part of the divergent integral for the retarded-field momentum in question. 
\end{abstract}


\maketitle

\noindent
Dondera \cite{don} has recently approached the problem of calculating the momentum of the retarded field of an arbitrarily moving point charge in a novel way. This momentum is given by a divergent integral
\begin{eqnarray}\label{mom}
{\bf G}(t)=\varepsilon_0\int d {\bf r}\; {\bf E}({\bf r},t) \times {\bf B}({\bf r},t),
\end{eqnarray}
where ${\bf E}$ and ${\bf B}$ are the well-known retarded Lienard-Wiechert fields, and by changing the integration variable in (\ref{mom}) to a retarded relative coordinate ${\bf R}= {\bf r}- {\bf r}_e(t_{\rm ret})$, where $ {\bf r}_e(t_{\rm ret})$ is the charge's position at the retarded time  $t_{\rm ret}=t-R/c$, Dondera was able to obtain for (\ref{mom})  an expression that is a sum of divergent and finite terms but does not involve any retarded quantities,
\begin{align}\label{gd}
{\bf G}_{\rm D}(t)  = &\lim_{\varepsilon \rightarrow 0} \frac{2e_0^2}{3c \,\varepsilon} \gamma^2 {\bm \beta}- 
\frac{2e_0^2}{3c^2} \frac{d}{dt}(\gamma^2 {\bm \beta}) \nonumber \\*
&+ \frac{2e_0^2}{3c^2} \int_{-\infty}^t dt' \gamma^4\left[ \dot{\beta}^2 + \gamma^2 ({\bm \beta}\cdot \dot{\bm \beta})^2\right] {\bm \beta}
\end{align}
(\cite{don},  the limit $\varepsilon {\rightarrow} 0$ of Eq.\,(14)). We use here Dondera's notation $e_0=e/\,\sqrt{4\pi \varepsilon_0}$; ${\bm \beta}={\bf v}(t)/c$, $\gamma =(1-\beta^2)^{-1/2}$ and the overdots indicate time derivatives.  Dondera aimed at deriving the space component of the relativistic LAD equation of motion of a point charge \cite{dir} using  the time derivative of ${\bf G}_{\rm D}(t)$, which can be obtained easily from (\ref{gd}).
However, ${\bf G}_{\rm D}(t)$ cannot be correct relativistically. Because of the factor $\gamma^2$ instead of $\gamma$, the divergent term of (\ref{gd}) cannot be absorbed in the  charge's momentum by a renormalization of its rest mass in a manner that is consistent with special relativity, while the negative of the time derivative of the sum of the finite terms agrees with the relativistic LAD radiation reaction force only in the leading term, $(2e_0^2/3c^2)\gamma^2 \ddot{{\bm \beta}}$. Dondera dealt with this ``defect" by endowing the accelerating point charge itself, which he calls the ``bare electron", with a momentum that ``compensate[s] both the singular and noncovariant terms" when it is added to the field momentum ${\bf G}_{\rm D}$ \cite{com}. 

In this Comment, we show that the finite and divergent terms of (\ref{gd}) result from the decomposition of the divergent integral (\ref{mom}) in which the integration variables are changed according to Dondera's transformation into the Hadamard finite part \cite{est1} and the corresponding divergent term. A change of the integration variables generally changes the Hadamard finite part \cite{est2} (reflecting the fact that a regularization of a divergent integral depends on the manner it implies of the approach to the singularity in the integrand \cite{hni2}), and we conjecture that the Hadamard decomposition of the divergent integral (\ref{mom}) with the original integration variables is 
\begin{align}\label{fp1}
{\bf G}(t)= \lim_{\varepsilon \rightarrow 0} \frac{2e_0^2}{3 c\,\varepsilon}\gamma {\bm \beta} + {\rm Fp} \, {\bf G}(t),
\end{align}
where 
\begin{align}\label{fp2}
{\rm Fp}\, {\bf G}(t)=-\int_{-\infty}^t dt'\, {\bf F}_{\rm LAD}(t')
\end{align}
is the Hadamard finite part with 
\begin{align}\label{fp3}
{\bf F}_{\rm LAD}(t) =& \frac{2e_0^2}{3c^2}\gamma^2 \left[ \ddot{\bm \beta} + 3\gamma^2 ({\bm \beta} \cdot \dot{\bm \beta})\dot{\bm \beta} \right.\nonumber \\*
& \left. + \gamma^2 ({\bm \beta} \cdot \ddot{\bm \beta}){\bm \beta} +3\gamma^4 ({\bm \beta} \cdot \dot{\bm \beta})^2 {\bm \beta} \right]
\end{align}
being the space component divided by $\gamma$ of the LAD radiation-reaction 4-force \cite{roh}. This conjecture challenges the cogency of endowing the ``bare electron" with attributes that would fix the ``defects" of the result of the calculation of a quantity that in principle should depend only on the particle's charge and trajectory. We shall support the conjecture by a recent calculation of the momentum of the retarded field of an accelerating point charge \cite{hni}.

The Hadamard finite part of a divergent 3-dimensional integral can be evaluated by using spherical coordinates and performing the angular integration first. When the integrand of the remaining radial integral can be written as $\Phi(r)/r^k$, where $\Phi(r)$ is a function regular at $r=0$, the finite part of the original integral is given by the finite part of the radial integral.
After implementing Dondera's change of variables and performing the angular integration, one obtains for the Cartesian components of (\ref{mom})
\begin{align}\label{6}
\frac{3c}{2e_0^2}G_i(t)=&\int_0^{\infty}dR \,\frac{f_i\left(t-\frac{R}{c}\right)}{cR} + \int_0^{\infty}dR \,\frac{g_i\left(t-\frac{R}{c}\right)}{R^2}\nonumber \\
& +\frac{1}{c^2}\int_0^{\infty}dR \, h_i\left(t-\frac{R}{c}\right), 
\end{align}
where 
\begin{align}
f_i(t-R/c)&=2\gamma^4({\bm \beta} \cdot \dot{\bm \beta}) \beta_i+ \gamma^2 \dot{\beta_i},\label{7}\\
g_i(t-R/c)&=\gamma^2 \beta_i,\\
h_i(t-R/c)&=\gamma^4 \left[ \dot{\beta}^2+ \gamma^2 ({\bm \beta} \cdot \dot{\bm \beta})^2 \right] \beta_i\label{9}
\end{align}
(\cite{don}, Eqs.\,(A9)--(A11)). The first two integrals in (\ref{6}) diverge and we evaluate their finite parts according to the formula 
\begin{align}\label{fp}
&{\rm Fp} \int_0^{\infty} dx \frac{\Phi(x)}{x^k} = \int_0^a  \frac{dx}{x^k} 
\left[\Phi(x)  
 -\sum_{j=0}^{k-1} \frac{\Phi^{(j)}(0)}{j!}x^j \right] \nonumber \\
 &+\int_a^{\infty} dx\frac{\Phi(x)}{x^k}-\sum_{j=0}^{k-2} \frac{\Phi^{(j)}(0)}{j!
 (k-j-1)a^{k-j-1}}\nonumber\\
 &+\frac{\Phi^{(k-1)}(0)}{(k-1)!}\ln a, 
\end{align} 
where $a$ is an arbitrary constant (\cite{est1}, Eq.\,(2.13)).
The corresponding divergent parts are then given by the limits $\epsilon {\rightarrow} 0$ of
\begin{align}\label{div}
\sum_{j=0}^{k-2} \frac{\Phi^{(j)}(0)}{j!(k-j-1)\epsilon^{k-j-1}}- \frac{\Phi^{(k-1)}(0)}{(k-1)!} \ln \epsilon
\end{align}
(note that here $\epsilon$ is dimensionless, unlike the $\varepsilon$ in (\ref{gd}) and (\ref{fp1})). It is convenient to use in (\ref{6}) a dimensionless variable of integration $x=R/R_0$, where $R_0$ is a fixed arbitrary length, and to choose $a=1$ in (\ref{fp}).
We use  the fact that $f_i(t)=dg_i(t)/dt$ and integrate term by term  the first integral on the 
r.h.s.\,of (\ref{fp}), after expanding the integrand in Taylor series. The infinite
series can be summed in terms of the functions $g_i$ and their time derivatives, and we obtain for the Hadamard finite part of (\ref{mom}) when Dondera's transformation is used the value
\begin{align}
{\rm Fp}\,{\bf G}(t)_{{\bf r}\to{\bf R}} =&- 
\frac{2e_0^2}{3c^2} \frac{d}{dt}(\gamma^2 {\bm \beta})\nonumber \\
&+ \frac{2e_0^2}{3c^2} \int_{-\infty}^t dt' \gamma^4\left[ \dot{\beta}^2 +\gamma^2 ({\bm \beta}\cdot \dot{\bm \beta})^2\right] {\bm \beta},
\end{align}
which equals the finite component of (\ref{gd}); the use of formula (\ref{div}) then yields the divergent part of (\ref{gd}).

To prove the conjecture (\ref{fp1})--(\ref{fp3}) by performing a calculation of the Hadamard finite part along the above lines seems well-nigh impossible because the integrand in (\ref{mom}) depends on the integration variable $\bf r$ not only explicitly but also implicitly through retardation. An exception is the case of a uniformly moving charge whose fields are expressible in terms of present-time quantities, and the Hadamard finite part of the resulting field momentum can be shown easily to vanish \cite{com2}.  While simplifying,  the case of non-uniform motion along a rectilinear trajectory retains the complexity due to retardation, and it is noteworthy that the evaluation in \cite{hni} of the divergent retarded integral (\ref{mom}) for this case has yielded a value equal to the finite part (\ref{fp2}), including the vanishing value when ${\bm\beta} = \rm{const}$,  with no need for any explicit removal of infinities. The  calculation procedure of \cite{hni} thus amounted to the extraction of a finite part of the requisite divergent integral.  The integration of the field-momentum density  was carried out there in the momentum space, using spherical coordinates of the integration variable. The Fourier transforms of the Lienard-Wiechert fields were calculated by integration by parts, employing distributional derivatives that discard the surface terms that arise in such integration when classical derivatives are used. Distribution theory shows that this is equivalent to the use of the Hadamard finite part for a divergent integral (\cite{kan}, Sec.\,4.1). 

Some 45 years ago, Rowe \cite{row} investigated an ``ambiguity" in the derivation of the LAD equation, reaching the conclusion that  the regularization effected by distributions disposes of the divergencies of standard classical electrodynamics.  His distribution-theoretic derivation of the LAD equation is free of renormalization and any other removal of infinities. We believe that the manifestly covariant derivation of Rowe, as well as the non-manifestly-covariant calculation of  \cite{hni}, lends strong support for our conjecture.
 
The divergent
 and noncovariant terms in the expression that Dondera obtained for the momentum of the retarded field of an accelerating point charge are not a defect to be repaired by endowing an accelerating ``bare" charge with {\it ad hoc} compensating attributes --- their presence is in fact mathematically well justified.  In standard classical electrodynamics, the divergent nature of these terms is due to the vanishing spatial extension assumed for the charge. The particular non-covariant features of the terms arise from the transformation of the integration variables that Dondera used in his evaluation of the pertinent retarded integral.


\begin{thebibliography}{30}
\bibitem{com} Dondera complemented in a similar way also the divergent noncovariant expressions he obtained for the energy and angular momentum of the charge's retarded field  using his transformation of the integration variables (to claim a derivation of the manifestly-covariant LAD equation of motion, he needed a relativistically correct finite result for the energy of the charge's field as well as that for the momentum).
\end{thebibliography}
\end{document}